%% file: main.tex
\documentclass{llncs}
\input{macros}

\title{Generative Interest Estimation\\For Document Recommendations}
\author{Danijar Hafner, Alexander Immer, Willi Raschkowski, Fabian Windheuser}
\institute{Hasso Plattner Institute, Potsdam, Germany}
\date{June 2016}

\begin{document}

\frontmatter
\maketitle

\begin{abstract}
\input{section/abstract}
\end{abstract}

\begin{keywords}
content-based filtering, recommendations, representation learning,\\density estimation, probabilistic generative models
\end{keywords}

\section{Introduction}
\input{section/introduction}

\section{Related Work}
\input{section/related}

\section{Generative Interest Estimation}
\input{section/algorithm}

\section{Experiments}
\input{section/experiments}

\section{Conclusion}
\input{section/conclusion}

\bibliography{references}

\end{document}

%% file: macros.tex
\usepackage[english]{babel}
\usepackage[utf8]{inputenc}
\usepackage[square,numbers]{natbib}
\usepackage{graphicx}
\usepackage{subfig}
\usepackage{authblk}
\usepackage{color}

\newcommand{\annotate}[1]{}

\newcommand{\todo}[1]{\annotate{TODO: #1}}

%% file: section/abstract.tex
Learning distributed representations of documents has pushed the state-of-the-art in several natural language processing tasks and was successfully applied to the field of recommender systems recently. In this paper, we propose a novel content-based recommender system based on learned representations and a generative model of user interest.

Our method works as follows: First, we learn representations on a corpus of text documents. Then, we capture a user's interest as a generative model in the space of the document representations. In particular, we model the distribution of interest for each user as a Gaussian mixture model (GMM). Recommendations can be obtained directly by sampling from a user's generative model.

Using Latent semantic analysis (LSA) as comparison, we compute and explore document representations on the Delicious bookmarks dataset, a standard benchmark for recommender systems. We then perform density estimation in both spaces and show that learned representations outperform LSA in terms of predictive performance.

%% file: section/introduction.tex
To many people, the internet has become the primary source of reading material. However, due to the vast amount of available content, it is a challenge to find interesting articles. Recommender systems for textual data address this problem. The two conceptual approaches to recommender systems are \emph{collaborative} and \emph{content-based filtering}.

In collaborative filtering, user attributes and information on his historic behavior are leveraged to make new recommendations. From a corpus of documents and the previously read documents of each user, underlying topic groups are inferred. Articles, read by a group of users, can be recommended to similar users.

While this is a common approach, it has downsides. Collaborative filtering suffers from the \emph{cold-start problem}. The relevance of documents can only be estimated if they were visited by other users. Thus, enough empiric records have to be collected first. These records need to contain enough overlap between user interests to effectively infer underlying interest groups that connect users and documents. 

In contrast, content-based filtering only requires preferred documents for each user. Documents are grouped semantically by their content so that no overlap in the users' preferences is required. The challenge in content-based filtering however, is to find useful semantic representations of documents.

\subsection{Vector representations of documents}

For content-based recommender systems, documents must be embedded into a common space. This space allows to relate documents to each other. There are several approaches to create semantic vector representations of text documents. We use learned representations and compare them with Latent Semantic Analysis (LSA) as described below.

A common document representation is bag-of-words (BOW)~\cite{zhang2010understanding} where documents are represented by their contained words. The vector of a document has the length of the overall vocabulary. The elements of this vector are one, if the corresponding word is contained in the document and zero otherwise. Note that this representation looses information about the ordering of words. BOW vectors are often reduced in dimensionality using SVD. Applying SVD to BOW features is known as LSA and further described in \cite{landauer1998introduction}. Other additions include weighting the elements of the vector using term frequency measures, such as tf-idf~\cite{sparck1972statistical}. However, the usefulness of those weightings in practice has been questioned~\cite{dai2015document}.

Recently, learning vector representations from large text corpora has become computationally feasible~\cite{mikolov2013distributed}. Learned representations have led to improvements in the state-of-the-art of natural language problems, such as part-of-speech tagging and named entity recognition~\cite{turian2010word, collobert2008unified}. \citet{le2014distributed} extended the idea in order to learn representations of phrases and documents. Those document representations are of particular interest to content-based recommender systems, since the learned vectors show promising semantic properties~\cite{dai2015document,musto1441word,kataria2015distributed}.

\subsection{Generative model of user interest}

Given fixed-length vector representations for the documents, there are multiple approaches to suggest new potentially relevant articles ~\cite{adomavicius2005toward}. A common approach for this is to compute a relevance score for each document unseen to the user. One can compute such a scoring using nearest neighbor queries as proposed in \citet{cover1967nearest}. For each unseen document, the $k$ nearest documents from the user documents are queried using a distance measure, such as cosine distance or euclidean distance. The score is composed of the $k$ computed distances. Limitations of this approach are the arbitrary choice of both $k$ and the distance function~\cite{bhatia2010survey}.

An alternative to using nearest neighbors is density estimation~\cite{sun2012estimating} using generative models~\cite{westerveld2007generative,hiemstra1998linguistically}. We model the interests of a user as a probability distribution over the semantic space. Assuming that the user's history was generated from his interest, we model it using a Gaussian mixture model (GMM). This allows to capture the user's interest being spread across multiple clusters in a semantic space. GMMs are a common choice for modelling densities and have already been used in the field of collaborative recommender systems and text modeling by \citet{hofmann2003collaborative} and \citet{hiemstra1998linguistically}, respectively. 

\subsection{Our contributions}

In this work, we suggest combining learned document representations with generative density estimation using GMMs. We focus on content-based recommendations in the domain of text documents. In Section 2, we will compare our approach to related work on recommendation systems that use learned representations and density estimation. Section 3 introduces our approach that we call density-based interest estimation. Following, we conduct a benchmark on the Delicious bookmark dataset and compare learned representations with BOW representations (Section 4). In Section 5, we summarize our findings and point to further directions of research.

%% file: section/related.tex
Content-based recommender systems form an active area of research. For an overview of state-of-the-art methods we refer to \citet{lops2011content} and \citet{adomavicius2005toward}. Our work differs from most of their systems in two ways. First, learned document representations are used instead of the common LSA representation, as described in \citet{mikolov2013distributed}. Second, we use density estimation to obtain a generative model of user interest.

Learned representations were only recently proposed for use in content-based recommender systems. \citet{musto1441word} study the difference between several document representations for content-based recommender systems. They suggest learned representations as a well-performing option compared to latent semantic indexing and random indexing. While they use word representations in their comparison, we use the closely related document representations in our work.

\citet{kataria2015distributed} present another approach using learned representations for content-based recommendations. They propose a two-step solution for recommending user tags using document embeddings and obtain remarkable results for tag recommendations. We focus on recommending documents instead of tags.

Building on the above research work, we suggest a generative approach that allows to recommend new documents by sampling from the estimated user interest rather than scoring all and sorting all candidate documents.

Based on learned document representations, we employ a generative model of user interest and estimate its density. \citet{westerveld2007generative} explain the usage of probabilistic models for classification tasks as used in our work.

Another common approach is the usage of a multinomial distribution with independency assumption as used in Naive Bayes classifiers~\cite{duda2012pattern}. However, these approaches do not use density estimation on learned representations, which are explored in this paper.

%% file: section/algorithm.tex
In general, our approach involves the three steps preprocessing (Section \ref{sec:preprocessing}), computation of document representations (Section \ref{sec:representation}), and density estimation (Section \ref{sec:estimation}). For the document vectors, we compare latent semantic analysis and learned representations.

\subsection{Preprocessing}
\label{sec:preprocessing}

Inputs to our method are a corpus of documents and a set of documents, representing the user's preference. The text of each document is preprocessed by removing special-characters and conversion to lower-case as described in \citet{srividhya2010evaluating}. Neither stemming is used, nor are stop words removed, to avoid introducing bias and removing potentially useful information that word classes contain. When computing document vectors, information captured in the endings of words will be discarded automatically, if not relevant for the task. From tokenized documents, we compute representations in two ways, as explained in the next two subsections.

\subsection{Latent semantic analysis}
\label{sec:lsa}

Latent Semantic Analysis (LSA) aims at representing words with a more general semantic structure. LSA involves transforming documents into a vector space model~\cite{salton1975vector}, followed by Latent Semantic Indexing (LSI), as first described by \citet{dumais2004latent}. \citet{deerwester1990indexing} claim that LSI can provide linguistic notion by capturing synonymy and polysemy. Their work shows that related semantic categories are close in the created vector space. LSA works in many cases because it groups words with similar co-occurrence patterns together~\cite{sahlgren2006word}.

\subsection{Learned representation}
\label{sec:representation}

The idea of learned word representations as proposed by \citet{mikolov2013distributed}, is to start with random vectors of a chosen fixed length. A shallow neural network is then trained on a large text corpus. The objective is to correctly predict the current word from its context, that is, the surrounding words. Using the backpropagation algorithm~\cite{hecht1989theory}, the gradient of the errors with respect to both the weights of the classifier and the input vector is calculated. We then update both the classifier and the word-vectors of the surrounding words using gradient descent. There are several modifications to this architecture. We suggest the papers \citet{mikolov2013efficient} and \citet{collobert2008unified} for comparison and further insights.

To learn distributed representations of documents, \citet{le2014distributed} extended the model as explained. The classifier is initialized with a random paragraph vector, in addition to the context words. This paragraph vector is trained via gradient descent, resulting in a paragraph vector that covers the overall semantics of a text. The authors suggest two versions of this algorithm. Either to only train the classifier and the paragraph vectors, or to train the word vectors as well. We only train the paragraph vectors because we found that training the word vectors does not improve performance significantly.

For a given set of context words, there are multiple plausible choices for the current word. We thus expect the paragraph vector to converge to the vector that best helps resolve this ambiguity. After training, each document has a trained vector representation in a semantic space, where similar documents are close to each other~\cite{dai2015document}. 

\subsection{Density estimation}
\label{sec:estimation}

Given fixed-length document representations, we now model the user interest. We assume that documents in a user profile $s_u$ were generated by some underlying distribution of user interest $p_u(d)$~\cite{lavrenko2008generative}.

We estimate the probability distribution assuming that each document $s_u$ is an i.i.d. realization of the users interest distribution $p_u(d)$. Further, we model this distribution as a weighted mixture of Gaussians. We fit the so called Gaussian mixture model to the $s_u$ using Expectation Maximization. The estimation takes place in the semantic space of document representations.

Using a mixture model, we capture the fact that users can be interested in multiple topics spread across the semantic space. This introduces the parameter $k$, which specifies the number of distributions. $k$ should be close to the amount of topics that the user is interested in. 

The quality to which the model can capture the user interest heavily depends on the semantic space and thus on the quality of document representations. Therefore, in the next section, we compare the density estimation with both, LSA and learned representations. 

%% file: section/experiments.tex
We conduct a benchmark on the Delicious dataset, common in literature on both collaborative and content-based recommender systems. We start with explaining the dataset and apply pre-processing steps. Afterwards, we explore the spaces produced by both LSA and learned representations on crawled documents of this dataset (Section \ref{compare_spaces}). Then, we explain our measure for evaluating the user interest models and compare the performance in both representation spaces (Section \ref{compare_models}).

The Delicious bookmarks dataset contains 105,000 bookmarks of more than 1,800 users~\cite{cantador2011workshop}. In order to embed the bookmarked pages, we crawl the corresponding websites and extract their main content by detecting the largest text area. In this step, we discard all bookmarks that are not accessible or have less than 500 characters of content, resulting in 46,000 documents. We then discard users with less than 50 bookmarked pages, following~\citet{cantador2010content}. The final dataset consists of 2,536 bookmarks from 50 different users. On average, a user holds 57 bookmarked documents.

\subsection{Comparison of document representations}
\label{compare_spaces}

Before evaluating the user interests' models, we explore the semantic spaces constructed by LSA and learned representations. For LSA, we use tf-idf weighted occurrence vectors and only consider terms with tf-idf scores within range $\left[0.3,10.0\right]$. After collecting the count vectors, we reduce their dimensionality to 10 components using singular value decomposition (SVD).

For learned representations, we use the paragraph2vec algorithm as described in \citet{mikolov2013distributed} with a vector size of $300$, $10$ training iterations, and $30$ noise-contrastive examples. We also reduce the dimensionality of the learned representations to $100$ components, using kernel PCA with RBF kernel.

\todo{why those parameters?}

\begin{figure}%
    \centering
    \subfloat[LSA representations]{{\includegraphics[width=5cm]{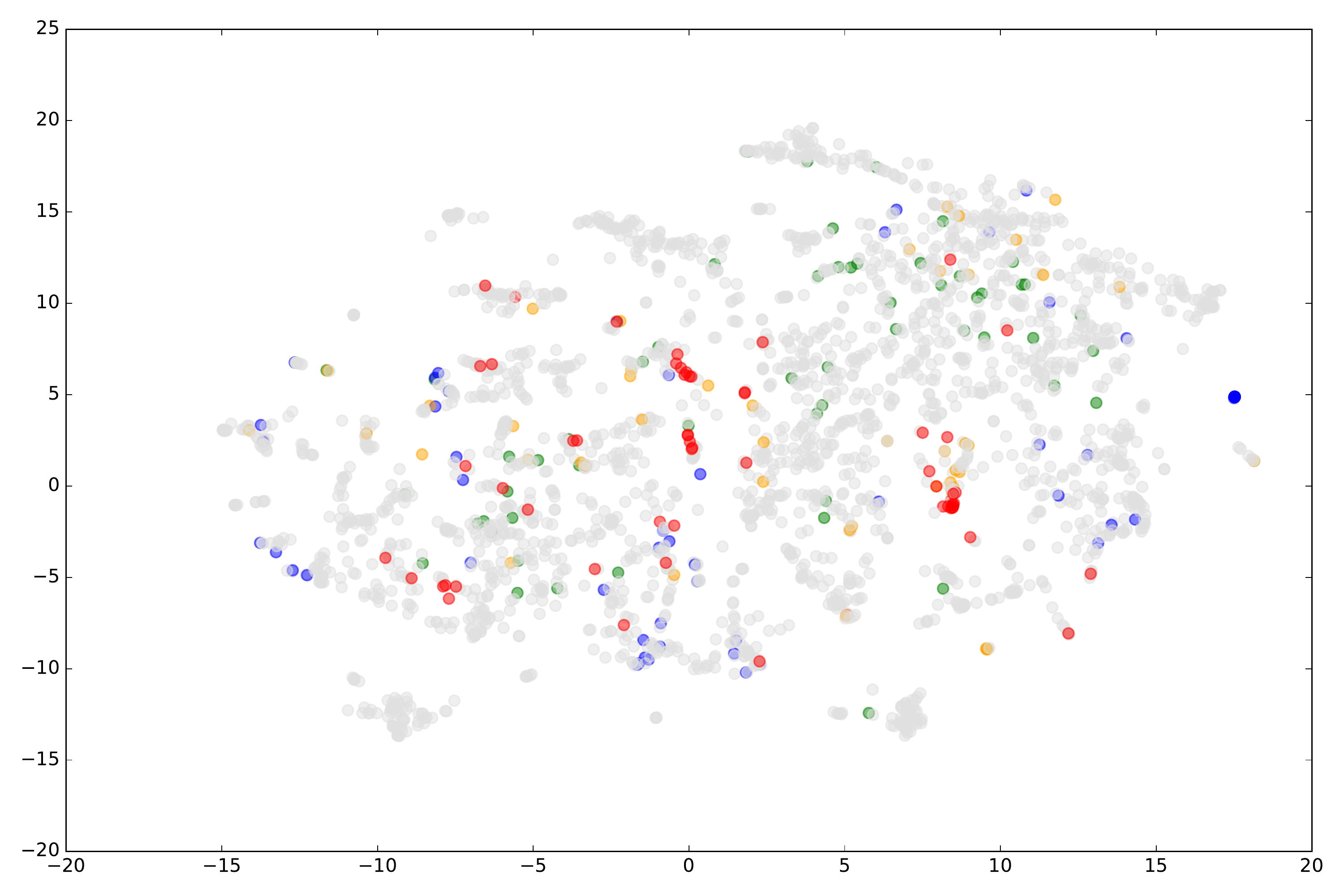} }}%
    \qquad
    \subfloat[Learned representations]{{\includegraphics[width=5cm]{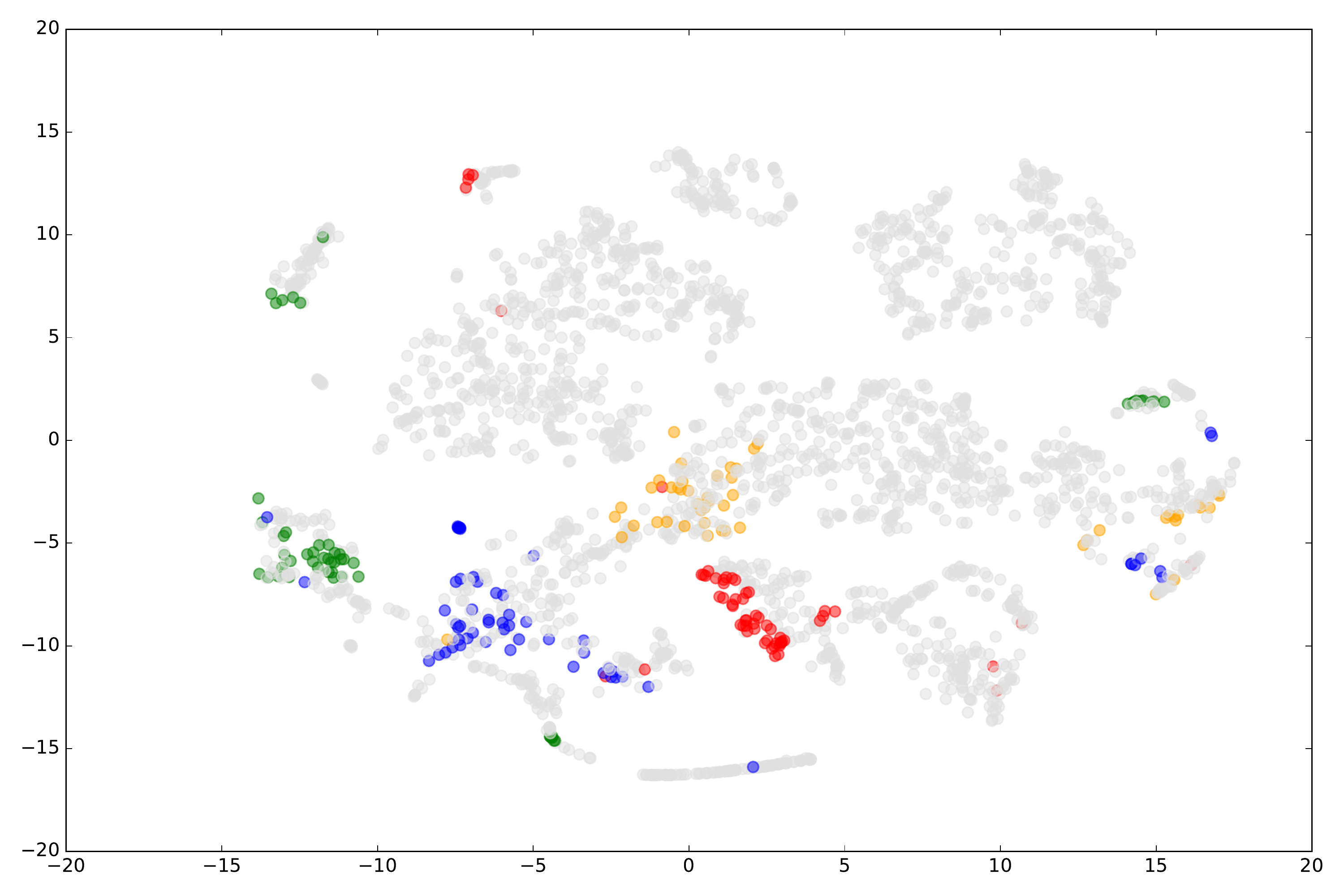} }}%
    \caption{Embedded corpus documents mapped into 2D space using t-SNE~\cite{maaten2008visualizing}. Four randomly chosen user profiles are highlighted. Without prior knowledge, documents of each user end up being grouped together.}%
    \label{fig:embeddings}%
\end{figure}

In Figure \ref{fig:embeddings}, we compare the learned vector spaces of both representations. The figure shows the corpus of all documents with four randomly selected user profiles being highlighted. While in LSA space, the documents of a user profile are spread over the map, learned representations seem to implicitly group users into one or more clusters. In the following section, we will show that this allows to model user interests more accurately.

\subsection{Evaluation of Density Models}
\label{compare_models}

While the goal of text-based recommender systems is to suggest useful documents, it is not obvious how to measure this formally. With a large empiric study, it would be possible to perform an evaluation based on user interaction with recommendations. However, such an evaluation is not feasible in many cases. Therefore, a common approach is to measure the quality based on the predictive performance of user behavior.

For each user, we conduct k-fold cross-validation with $k=5$ to split the user profile into training data $D_t$ and validation data $D_v$. For each split, we fit a GMM to $D_t$ and sample $|D_v|$ times from the model. Since the samples lie in continuous space, we pick the nearest corpus document for each sample, measured by euclidean distance. Finally, we count the number of sampled documents that match documents of $D_v$ as hits.

The hit rate over all splits and users is a measure of predictive performance for the user interests. Each user profile just contains samples from the underlying distribution so that we can expect other corpus documents even in high-density areas of a user interest. Therefore, we expect the hit rates to be vastly below 100\%.

Results are shown in \ref{results}. Estimating user interests from learned representations significantly outperforms using LSA representations. This confirms our assumption from the previous section, that learned representations group documents in a way that helps capturing user interests.

\begin{table}
\begin{center}
\setlength{\tabcolsep}{0.5em}
{\renewcommand{\arraystretch}{1.2}
\begin{tabular}{lrr}
  \hline
  Method & Hit rate & Hits \\
  \hline \hline
  GMM on LSA representations & 8.63\,\% & 244 / 2828 \\
  GMM on learned representations & 12.80\,\% & 362 / 2828 \\
  \hline
\end{tabular}}
\\[10pt]
\caption{Performance of the generative user interest model in LSA or learned representation space. The error is calculated as the average number of test documents hit by sampling from the respective user interest. Our proposed method significantly improves performance over the common LSA approach.}
\label{results}
\end{center}
\end{table}

%% file: section/conclusion.tex
We proposed a generative approach to modelling user interests based on user profiles of preferred documents. Based on the recently introduced method of learned document representations, our model effectively captures user interest. We show this by comparison with the established approach of LSA, on a benchmark common in the literature on recommender systems. We conclude that learned representations create effective semantics for content-based recommendations. Further exploration is needed to fully understand the reasons for this effectiveness. The areas in which learned representations or established methods are preferable are another worthwhile direction for further research.